\documentclass[a4paper,fleqn,usenatbib,useAMS]{mnras}

\usepackage{graphicx}	
\usepackage{amsmath}	
\usepackage{amssymb}	
\usepackage{multicol}        
\usepackage{bm}		
\usepackage{pdflscape}	

\usepackage[T1]{fontenc}
\usepackage{ae,aecompl}

\usepackage{txfonts}

\title[An IR-dark cloud hosting an IMBH candidate]{\textit{La Freccia Rossa}: An IR-dark cloud hosting the Milky Way intermediate-mass black hole candidate} 

\author[ Ravi, Vedantham \& Phinney ]{
Vikram Ravi$^{1}$\thanks{Contact e-mail:vikram@caltech.edu},
Harish Vedantham$^{1}$, 
E. Sterl Phinney$^{2}$\\
$^{1}$Cahill Center for Astronomy and Astrophysics MC 249-17, California Institute of Technology, Pasadena CA 91125, USA. \\
$^{2}$Theoretical Astrophysics, MC 350-17, California Institute of Technology, Pasadena CA 91125, USA.
}

\date{2017 October 10}

\pubyear{2017}

\begin{document}
\label{firstpage}
\pagerange{\pageref{firstpage}--\pageref{lastpage}}
\maketitle

\begin{abstract}

The dynamics of the high-velocity compact molecular cloud CO-0.40-0.22 have been interpreted as evidence for a $\sim10^{5}M_{\odot}$ black hole within 60\,pc of Sgr A*. Recently, Oka et al. have identified a compact millimetre-continuum source, CO-0.40-0.22*, with this candidate black hole. Here we present a collation of radio and infrared data at this location. ATCA constraints on the radio spectrum, and the detection of a mid-infrared counterpart, are in tension with an Sgr A*-like model for CO-0.40-0.22* despite the comparable bolometric to Eddington luminosity ratios under the IMBH interpretation. A protostellar-disk scenario is, however, tenable. CO-0.40-0.22(*) is associated with an arrowhead-shaped infrared-dark cloud (which we call the {\em Freccia Rossa}). Radio-continuum observations reveal a candidate HII region associated with the system. If the $V_{\rm LSR}\approx70$\,km\,s$^{-1}$ systemic velocity of CO-0.40-0.22 is common to the entire Freccia Rossa system, we hypothesise that it is the remnant of a high-velocity cloud that has plunged into the Milky Way from the Galactic halo. 

\end{abstract}




\section{Introduction}

\citet{oti+17} present evidence for a $\sim10^{5}M_{\odot}$ intermediate-mass black hole (IMBH) lurking within the central 60\,pc of the Milky Way. The first hint of this object was the discovery of a compact molecular cloud with an unusually high line-of-sight velocity, and velocity dispersion \citep[C0-0.40-0.22;][]{oon+12,tom+14}. Such clouds \citep{onk+07} are common within the Central Molecular Zone (CMZ) of the Galaxy, and have few-parsec sizes and velocity spreads of a few tens of km\,s$^{-1}$. Many of these clouds lack unambiguous sources of the high internal velocities, 
and their origins have been attributed to shocks caused by unseen supernovae \citep{onk+07}, and turbulence in colliding clouds or expanding molecular shells \citep{tom+14}. However, with an internal kinetic energy of $10^{49.7}$\,erg \citep{omm+16} and its lack of an expanding velocity structure, C0-0.40-0.22 appears inconsistent with these explanations. Its dynamics are consistent with a slingshot from an unseen $\sim10^{5}M_{\odot}$ compact object \citep{omm+16}. ALMA images with $\sim1$\arcsec~ resolution at 231\,GHz and 266\,GHz revealed a point-like continuum source within C0-0.40-0.22 \citep{oti+17}. This source (C0-0.40-0.22*) is interpreted by \citet{oti+17} as an analogue of the Galactic-centre black hole Sgr~A*. 

IMBHs with masses between $10^{2}-10^{6}M_{\odot}$ are predicted to exist in the present-day Universe as a general consequence of supermassive black hole seed formation \citep{v10,g12}. 
The likely numerous historical minor mergers of the Milky Way with companion dwarf galaxies \citep{gwn02,vgo+03} may plausibly have led to an IMBH sinking to the center of the Milky Way potential well \citep{c14}. If C0-0.40-0.22* is such an IMBH, it would provide an important clue to the merger history of the Milky Way, and provide a new site for the study of the interaction between massive black holes and their environments. 

\vspace{-0.03cm}

Thus motivated, we use the wealth of multi-wavelength data available at the position of C0-0.40-0.22* to better characterise this candidate IMBH and its environment. We collated archival 24.6\,GHz continuum observations from the Australia Telescope Compact Array (ATCA), mid-IR observations from the {\em Spitzer} space telescope, and Ks-band observations from the Visible and Infrared Survey Telescope for Astronomy (VISTA), and augmented these with new continuum observations from the ATCA. In \S2, we present and analyse measurements of the spectral energy distribution (SED) of C0-0.40-0.22*. Then, in \S3, we characterise the environment of C0-0.40-0.22*. We discuss our results in \S4, and conclude in \S5. 

\section{The ALMA point-source CO-0.40-0.22*}


\citet{oti+17} detected C0-0.40-0.22* in ALMA continuum observations as an unresolved source (synthesised beam of 1.35\arcsec$\times$0.55\arcsec), with flux densities of $8.38\pm0.34$\,mJy at 231\,GHz, and $9.91\pm0.74$\,mJy at 266\,GHz. The implied spectral index was $\alpha=1.18\pm0.65$.\footnote{The spectral index, $\alpha$, is defined for a flux-density spectrum $F_{\nu}\propto\nu^{\alpha}$.} \citet{oti+17} also presented a $3\sigma$ upper limit on the $1-7$\,keV flux of $1.4\times10^{-14}$\,erg\,s$^{-1}$\,cm$^{-2}$. 

We obtained centimetre-wavelength observations of CO-0.40-0.22* with the ATCA on 2017 September 8 (13:50\,UT). The six 22\,m dishes of the array were arranged in the H168 configuration, with five dishes in a compact `T' with a maximum baseline of 192\,m, and the sixth dish located approx. $4400$\,m from the `T'. Full-Stokes data were recorded with the Compact Array Broadband Backend \citep[CABB;][]{wfa+11} in a standard {\em 64M-32k} continuum setup, in sidebands centred on 32.5\,GHz and 36\,GHz. The flux density scale was set using a 10\,min observation of PKS\,1934$-$638, and the receiver bandpasses and complex gains were calibrated using the quasar PKS\,1741$-$312 (1.9$^{\circ}$ distant from the program source). Observing conditions were excellent, with negligible wind and $<100\,\mu$m of rms atmospheric path-length variation.  We were able to track the atmospheric phase variations on our longest baselines with sufficient accuracy by cycling between 2\,min observations of CO-0.40-0.22* and 40\,s observations of PKS\,1741$-$312. The pointing of each antenna was checked and corrected every 30\,min. The total time spent observing CO-0.40-0.22* was 61\,min.  

We reduced and calibrated our data using standard techniques with the MIRIAD software \citep{stw95}. To search for unresolved emission from CO-0.40-0.22*, we made a multi-frequency synthesis image of our data from both sidebands simultaneously with uniform weighting. The full-width half-maximum (FWHM) of the synthesised beam was $0.99\arcsec\times0.23\arcsec$. No source was detected at the position of CO-0.40-0.22*; we set a $3\sigma$ upper limit on its 34.25\,GHz flux density of 0.285\,mJy. 

We also analysed archival mid-IR data from the {\em Spitzer} IR Array Camera (IRAC) obtained as part of the Galactic Legacy Infrared Mid-Plane Survey Extraordinaire (GLIMPSE) program \citep{cbm+09}. We downloaded calibrated image cutouts in all four IRAC bands (3.6\,$\mu$m, 4.5\,$\mu$m, 5.8\,$\mu$m, 8.0\,$\mu$m) from the IRSA Cutouts Service\footnote{http://irsa.ipac.caltech.edu/applications/Cutouts/spitzer.html} at the position of CO-0.40-0.22*, rendered with the original 1.2\arcsec~pixels. A faint point-like source was evident at the position of CO-0.40-0.22* in the 3.6\,$\mu$m and 4.5\,$\mu$m images. As the field was exceedingly crowded, complicated by diffuse emission, and because the source was 9\arcsec~from a brighter star, we estimated the flux density of the source by summing the signal within a 1-pixel radius aperture centered on the position of CO-0.40-0.22*, and subtracted the mean background level from a 1-pixel annulus surrounding the aperture. The resulting flux densities were $0.16\pm0.07$\,mJy at 3.6\,$\mu$m, and $0.21\pm0.09$\,mJy at 4.5\,$\mu$m. The small aperture, in combination with the unknown interstellar extinction towards CO-0.40-0.22*, implies that these measurements are lower limits. 

\begin{figure}
    \centering
    \includegraphics[angle=-90,scale=0.65]{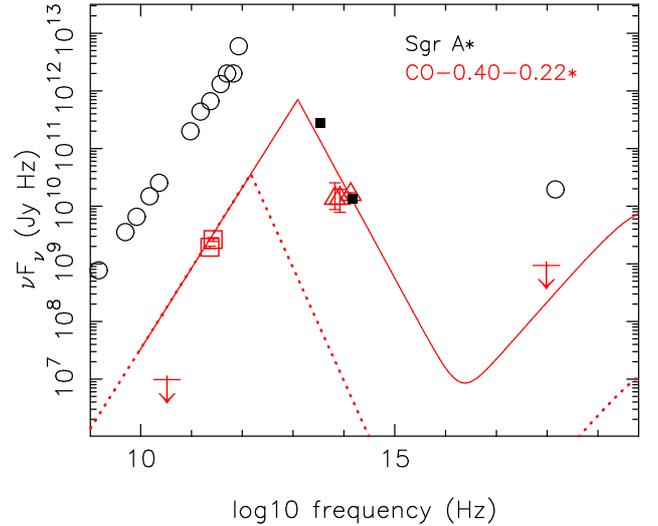}
    \caption{The broadband SED of the IMBH candidate CO-0.40-0.22* (red arrows, open squares, triangles), in comparison with the quiescent SED of Sgr A* (black circles, solid squares). The 231\,GHz and 266\,GHz detections of CO-0.40-0.22* (red open squares) and the X-ray $3\sigma$ upper limit (red arrow) were presented by \citet{oti+17}. The 34.25\,GHz $3\sigma$ upper limit (red arrow) and the IR detections (red triangles) were obtained through the analysis presented herein. The quiescent SED of Sgr A* was compiled from the references listed in Figure 1 of \citet{geg10}. Detections of Sgr A* are shown as open black circles, and upper limits are indicated by solid black squares. Neither SED has been corrected for extinction in the IR bands, or for interstellar absorption in the X-ray bands. Model ADAF SEDs for CO-0.40-0.22*, with $m=10^{5}M_{\odot}$ and $m=2.5\times10^{4}M_{\odot}$, are shown as dotted and solid curves respectively.}
    \label{fig:1}
\end{figure}

This mid-IR source was also detected in the Ks-band in a deep stack of images from the VISTA Variables in the Via Lactea survey \citep{mle+10}. We obtained pipeline-processed, calibrated and stacked data from the VISTA Science Archive \citep{ccm+12}, and performed standard aperture photometry. The measured flux density was $0.17\pm0.08$\,mJy. The unknown extinction again implies that this is a lower limit. The full SED of CO-0.40-0.22* is presented in Fig.~\ref{fig:1}. For comparison, we show the SED of the Galactic-Centre black hole Sgr A* compiled as described in Figure~1 of \citet{geg10}. 

\subsection{A synchrotron interpretation}

Assuming the distance of Sgr A* \citep[8.3\,kpc;][]{gef+13}, the bolometric luminosity of CO-0.40-0.22* is $\sim10^{35}$\,erg\,s$^{-1}$, or eight orders of magnitude less than the Eddington luminosity of a $10^{5}M_{\odot}$ IMBH. This ratio is comparable to that of Sgr A* \citep{geg10}. We therefore first consider whether the SED of CO-0.40-0.22* can be explained through mechanisms analogous to those attributed to Sgr A*.\footnote{We assume that the CO-0.40-0.22* SED is not affected by a variable accretion rate between epochs.} In its quiescent state, the peak of the Sgr A* SED can be largely explained by cyclo-synchrotron radiation \citep[e.g.,][]{opn00,wm06} from thermal ($T_{e}\sim10^{10}$\,K) electrons in an advection dominated accretion flow \citep[ADAF; e.g.,][]{rbb+82,m97,yqn03}. Some ultraviolet and X-ray flux is expected due to inverse-Compton (IC) scattering of the synchrotron emission. If CO-0.40-0.22* is an analogue of Sgr A*, the mid-IR detections imply that its spectral peak must be at approximately an order-of-magnitude higher frequency than Sgr A*. Furthermore, the $32-266$\,GHz spectrum of CO-0.40-0.22* is harder than that of Sgr A*. 

We attempted to fit an ADAF emission model to the SED of CO-0.40-0.22* using the formulation of \citet{m97}. We assumed standard ADAF parameters, specified by the viscosity parameter ($\alpha=0.3$), the magnetic-to-gas pressure ratio ($\beta=0.5$), the fraction of viscous heating transferred to electrons ($\delta=5\times10^{-4}$), and the minimum (3 gravitational radii $r_{g}$) and maximum ($r_{\rm max}=10^{3}r_{g}$) radial distances from the black hole with a constant $T_{e}$. The free parameters include the black-hole mass, $m$, and the accretion rate in units of the Eddington rate, $\dot{m}$; an energy-balance condition fixes $T_{e}$ for specific values of $m$ and $\dot{m}$. Our arguments are robust to these assumptions. We evaluate the optically thick synchrotron spectrum from the minimum frequency $\nu_{\rm min}$ (set by $r_{\rm max}$) to the peak frequency; this consists of the superposition of self-absorbed thermal cyclo-synchrotron peaks emitted at lower frequencies at larger radii. We then evaluate the IC and thermal brehmsstrahlung emission above the SED peak. 

For a given $m$, $\dot{m}$ is fixed by the ALMA detections of CO-0.40-0.22*. For $m=10^{5}M_{\odot}$, we find $\dot{m}=5\times10^{-7}$ and $T_{e}=9.2\times10^{9}$\,K. However, as shown by the dotted curve in Fig.~\ref{fig:1}, the resulting spectral peak is far too low in frequency to explain the IR emission, and the spectrum is also inconsistent with the 34\,GHz upper limit. If we vary $m$, we find a consistent solution with $m=2.5\times10^{4}M_{\odot}$, $\dot{m}=3.5\times10^{-5}$ and $T_{e}=4.8\times10^{9}$\,K (solid curve in Fig.~\ref{fig:1}); this solution is not unique. The resulting $\dot{m}$ is only a factor of few below the Bondi-Hoyle accretion rate for this system. 
The tension between our 34\,GHz limit and this model could be removed by reducing the ADAF's $r_{\rm max}$ by a factor of 3, removing the outer torus's dominant contributions to the $F_\nu\propto \nu^{2/5}$ spectrum, and leaving only the steeper ($\alpha\simeq 22/13$) quasi-black body from beyond $r_{\rm max}$. However, $r_{\rm max}\sim300r_{g}$ is unnaturally low for the equalisation of electron and proton temperatures to occur in standard two-temperature ADAF models at such low accretion rates.

On the other hand, the SED of CO-0.40-0.22* is also consistent with non-thermal synchrotron emission from a relativistic jet or wind. The flat 231\,GHz to IR spectrum is consistent with the $F_\nu\propto \nu^{1/3}$ optically thin synchrotron emission from relativistic electrons that are monoenergetic, or whose energy distribution has a sharp lower energy cutoff. The characteristic Lorentz factors $\gamma$ can be estimated from the fact that the synchrotron radiation is emitted at a frequency $\gamma^{2}\nu_{B}$, where $\nu_{B}$ is the cyclotron frequency. For a synchrotron peak in the IR, at $\nu=10^{14}$\,Hz, we have $\gamma^{2}B_{\rm kG}=3.6\times10^{4}$, where $B=B_{\rm kG}$\,kG is the magnetic field in the emission region. An upper limit on $B$ can be placed by constraining the emission at 231\,GHz to be optically thin. Then, $B\lesssim250R_{11}^{4}$\,G, where $R=10^{11}R_{11}$\,cm is the source radius, and $\gamma\gtrsim400R_{11}^{-2}$. Optically thin emission implies a brightness temperature below the equipartition value of $\sim10^{11}$\,K \citep{r94}, which in turn implies $R_{11}\gtrsim1=7r_{g}$, where $r_{g}$ is specified for a $10^{5}M_{\odot}$ IMBH. The 34\,GHz upper limit implies that the SED must transition to optically thick emission between 34\,GHz and 231\,GHz, implying $R_{11}\lesssim1.3$ at 34\,GHz. Depending on the transition frequency between optically thick and thin emission, and the intervening absorbing column, detectable IC radiation may be expected in the far-ultraviolet to X-ray band. We encourage a search for such emission.


\subsection{A thermal interpretation}

The alternative hypothesis for CO-0.40-0.22* is that the radio emission detected by \citet{oti+17} is thermal in nature. The best possibility is blackbody emission from warm dust. A hyper-compact HII region \citep{k05} is unlikely, because the $32-266$\,GHz spectral index constraint would imply optically thick emission at 266\,GHz, which with a canonical HII region brightness temperature of 8000\,K implies an amazingly small angular size of 5\,mas (40\,AU at Sgr A*). Additionally, hot molecular cores \citep{kcc00} and cold cores \citep{rjs06} are not viable interpretations for CO-0.40-0.22*. Hot cores are often associated with H$_{2}$O maser emission and molecular outflows, neither of which are observed in CO-0.40-0.22* \citep[SIMBAD;][]{oti+17}, and the size of CO-0.40-0.22* ($\lesssim0.04\times[D/(8.3\,{\rm kpc})]$\,pc from the ALMA observations) is smaller than expected for cold cores.


Warm ($\sim20-60$\,K) dust grains in protostellar disks can result in $30-300$\,GHz spectral indices in the range $\alpha\sim 2-2.5$ \citep{sgp+10}. In this scenario, the ALMA-detected flux density can be used to estimate the mass of the circumstellar material (dust and gas). Following \citet{bsc+90}, temperatures in the range $\sim20-60$\,K give disk masses in the range $1.2-4.4M_{\odot}\times[D/(8.3\,{\rm kpc})]^{-2}$. Such disks are common around young massive stars \citep[e.g.,][]{jtf+05}. 
The implied UV continuum from the central stellar object may create an HII region that dominates the radio spectrum at lower frequencies \citep{zrh+06}.
The dust disk would be transparent to IR emission, implying that the IR detections of CO-0.40-0.22* may be due to the central stellar object. Observations at frequencies $>266$\,GHz should reveal a spectrum that rises up to the dust thermal peak at a few terahertz, and observations at $<32$\,GHz may reveal an HII region. 


\section{A coincident IR-dark cloud}

\begin{figure*}
    \centering
    \includegraphics[trim={5.0cm 3cm 5cm 1.5cm},clip,scale=0.75,angle=90]{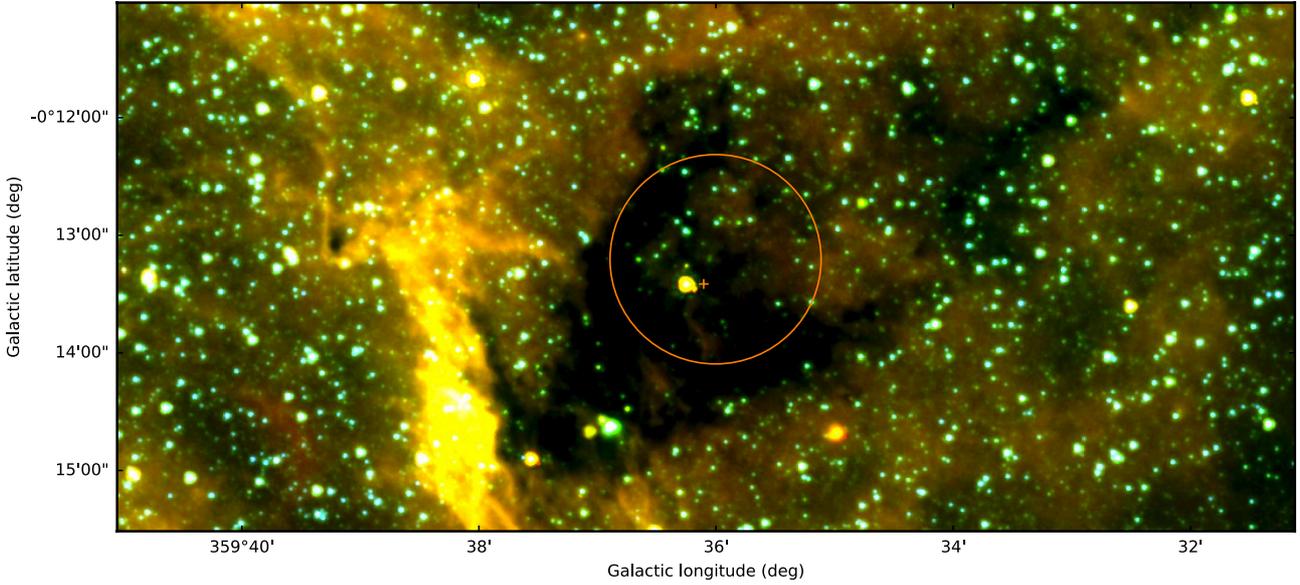}
    \caption{Three-colour mid-IR image of the {\em Freccia Rossa} IR-dark cloud from {\em Spitzer} GLIMPSE data. The red, green and blue components correspond to channels 4 (8\,$\mu$m), 3 (5.8\,$\mu$m) and 1 (3.6\,$\mu$m) of the IRAC instrument. The cross indicates the position of the IMBH candidate CO-0.40-0.22*, and the circle indicates the position and extent of the CO-0.40-0.22 high-velocity compact cloud \citep{tom+14,oti+17}.}
    \label{fig:2}
\end{figure*}

In our inspection of {\em Spitzer}/IRAC images at the position of CO-0.40-0.22*, we found an arrowhead-shaped region of high extinction against the diffuse Galactic mid-IR background and stellar field. This region (Fig.~\ref{fig:2}), which we term the Freccia Rossa, is coincident in position with CO-0.40-0.22*, and is additionally comparable in position and extent to the high-velocity compact molecular cloud CO-0.40-0.22. The Freccia Rossa has previously been identified as an IR-dark cloud (IRDC) in near-IR \citep[DC\,6;][]{nsn+09} and mid-IR data \citep[MSXDC\,G359.60$-$0.22;][]{sjr+06}. Based on an analysis of the near-IR stellar field obscured by the cloud, \citet{nsn+09} suggest a distance of 3.6\,kpc from the Earth, placing it within a chain of IRDCs possibly in the Norma spiral arm. 

The Freccia Rossa is also detected in (thermal) emission in the 1.1\,mm continuum Bolocam Galactic Plane Survey \citep[BGPS\,G359.62$–$0.24;][]{bab+10} with a comparable morphology to the extinction nebula. The inferred dust mass is $3\times10^{4}M_{\odot}$, which implies a gas mass of $\sim3\times10^{6}M_{\odot}$ and confirms this as a giant molecular cloud. The inferred gas column density of $N_{H_{2}}=1.2\times10^{24}$\,cm$^{-2}$ implies a CO $J=1-0$ line intensity of $\approx6\times10^{3}$\,K\,km\,s$^{-1}$ \citep{bwl13}, which is high, but consistent with the measured intensity of CO-0.40-0.22 \citep{ohs+98}. It is therefore likely that CO-0.40-0.22 is physically associated with the Freccia Rossa, as there is no other concentration of molecular gas along the line of sight \citep{ohs+98,tom+14}. Additionally, \citet{bab+10} contend that the cloud is in fact within the CMZ, rather than in the Norma arm, based on statistical arguments. In either case, the physical association of CO-0.40-0.22 and the Freccia Rossa implies a surprisingly large velocity of $V_{\rm LSR}\approx-70$\,km\,s$^{-1}$ for the system \citep{ohs+98,tom+14}. This velocity is likely systemic relative to the Galactic rotation because of the Galactic-centre sightline. 

\begin{figure}
    \centering
    \includegraphics[scale=0.38,angle=-90]{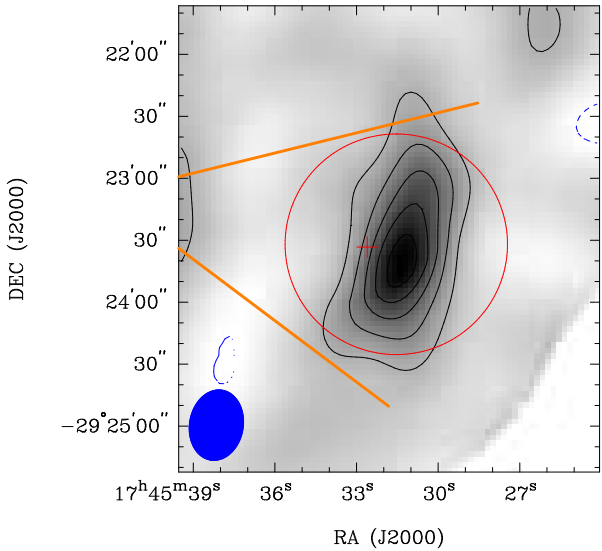}
    \caption{ATCA radio-continuum mosaic image of the CO-0.40-0.22 molecular cloud at a frequency of 24.6\,GHz. The contours are at $-$1.5 (dashed blue), 1.5, 3.0, 4.5, 6.0 and 7.5 mJy (solid black). The red cross indicates the position of CO-0.40-0.22*, and the red circle indicates the position and extent of CO-0.40-0.22 \citep{tom+14,oti+17}. The synthesised beam FWHM is indicated by a solid blue ellipse in the bottom-left corner. The orange wedge illustrates the approximate position of the Freccia Rossa cloud.}
    \label{fig:3}
\end{figure}

We searched our ATCA observations for emission on larger angular scales, which could be associated with the Freccia Rossa system. Internal heating of the CO-0.40-0.22 cloud above the extinction-nebula temperature can result in an HII region, detectable in the radio continuum. Upon making an image of our data in both the 32.5\,GHz and 36\,GHz sidebands with a $uv$ taper corresponding to an image-plane Gaussian of 35\arcsec~FWHM ($1/3$ of the primary beam FWHM) and a robust parameter of 0.5, a resolved source of $1.5\pm0.2$\,mJy peak flux density was detected at a position coincident with CO-0.40-0.22. This source was however poorly characterized by our observations, as only a few baselines were sensitive to this angular scale. We therefore searched the ATCA archive for observations at this location, and found 24.6\,GHz continuum data obtained with an identical CABB setup, but in the more compact H75 configuration, as part of the C2927 project. Three pointings of a mosaic, totalling 580\,s of integration time, were observed near the position of CO-0.40-0.22. We calibrated and imaged each pointing as described above, except that the long baselines to antenna 6 were excluded as they were not possible to calibrate. The three images were then summed with primary-beam weighting to produce a mosaic image at the position of CO-0.40-0.22 (Fig.~\ref{fig:3}). 

An extended source with an FWHM of $40\arcsec\times80\arcsec$ is detected at a position coincident with CO-0.40-0.22, with a peak flux density of $8.3\pm0.4$\,mJy. 
The spectral index of this emission between these 24.6\,GHz data and the 34.25\,GHz detection is difficult to estimate, because much of the emission is on angular scales larger than are probed by the 34.25\,GHz data. The source is nonetheless likely to be thermal brehmsstrahlung emission from an HII region, as it is too bright to correspond to the dust blackbody emission detected at 1.1\,mm \citep{bab+10}. Synchrotron emission from, for example, a supernova remnant is unlikely given the association with and comparable angular scales of the emission and the molecular cloud. In addition, no coincident emission is observed at 150\,MHz with a comparable angular resolution in the the TIFR GMRT Sky Survey \citep[TGSS;][]{ijm+17}, with a $3\sigma$ upper limit of 45\,mJy. This implies a 150\,MHz -- 24.6\,GHz spectral index of $\alpha>-0.33$, which is flatter than most radio supernova remnants \citep{dg15}. This is, however, consistent with an HII region.

\section{Discussion}

We first address the arguments made by \citet{omm+16} and \citet{oti+17} for the presence of an IMBH associated with the CO-0.40-0.22 cloud. Our data do not exclude a synchrotron origin for the ALMA-detected emission from CO-0.40-0.22*, but tension exists with Sgr A*-like models. 
However, the presence of an IR counterpart, and our lower limit on the radio spectral index, makes a protostellar-disk interpretation of CO-0.40-0.22* plausible. 

If CO-0.40-0.22* is not an IMBH, what could cause the large spread of velocities in CO-0.40-0.22? Of the explanations considered by \citet{tom+14}, the possibility of an expanding molecular shell driven by multiple supernovae is least impacted by our work. CO-0.40-0.22 lies on the edge of a large molecular shell \citep[shell 1 of][]{tom+14}. Bipolar outflows driven by deeply embedded stars may be indicated by the coincidence of CO-0.40-0.22 and the radio-detected HII region. However, if the large systemic velocity of CO-0.40-0.22 relative to the Galactic rotation matches the systemic velocity of the Freccia Rossa cloud, none of the mechanisms suggested by \citet{tom+14} are feasible. This is because the total mass of $\sim3\times10^{6}M_{\odot}$ and the systemic velocity of 70\,km\,s$^{-1}$ together imply an exceedingly large kinetic energy of $\sim10^{53}$\,erg. 

\subsection{A plunging halo object?}

A compelling hypothesis for the Freccia Rossa is that of a high velocity cloud (HVC) from the Galactic halo plunging into the Milky Way disk. HVCs detected in HI line emission, with masses $>10^{6}M_{\odot}$ and  velocities $V_{\rm LSR}>90$\,km\,s$^{-1}$, have long been observed to populate the Milky Way halo \citep{ww97}. Cold ($\sim10$\,K) dust has also recently been associated with HVCs \citep{mbr+05}. Although little evidence exists for HVCs that have impacted the Milky Way disk, simulations \citep[e.g.,][]{kh09} suggest that clouds with masses $\gtrsim3\times10^{5}M_{\odot}$ should survive the fall through the inner halo of the Galaxy. Recently, \citet{pkk+16} associated an HVC with a supershell of HI in the outskirts of the Galactic disk.

Under this hypothesis, the dynamics of the molecular cloud CO-0.40-0.22, and its high systemic velocity that we associate with the Freccia Rossa, are explained by the bulk motion of the system and the ensuing interaction with the Milky Way interstellar medium. The cometary morphology of the Freccia Rossa, with the densest dust to be found at the tip of the cloud \citep{bab+10}, provides circumstantial evidence for this scenario. The candidate HII region associated with CO-0.40-0.22 indicates ongoing star formation. We then have no extrinsic reason to interpret the compact source CO-0.40-0.22* as an IMBH; a protostellar disk is an alternative possibility, and consistent with the star-formation activity. Detailed imaging and modelling of the dust continuum and molecular-line emission will help unravel the dynamics of the Freccia Rossa system.

\section{Conclusions}

We examine the evidence for an IMBH interpretation of the high-velocity compact molecular cloud CO-0.40-0.22 \citep{omm+16}, and the associated millimetre-continuum source CO-0.40-0.22* \citep{oti+17}. The broadband SED of CO-0.40-0.22*, with its steep ($\alpha>1.77$) radio spectrum and IR counterpart, is in tension with Sgr A*-like models despite the comparably under-luminous nature assuming a $10^{5}M_{\odot}$ accreting object. Although we cannot exclude a synchrotron origin for the emission, we show that the system is also consistent with emission from a protostar surrounded by a $1.2-4.4M_{\odot}\times[D/(8.3\,{\rm kpc})]^{-2}$ disk. Further broadband measurements of the SED will establish the nature of CO-0.40-0.22*.

We also find that CO-0.40-0.22(*) is associated with an arrowhead-like IR-dark cloud with a total mass of $\sim3\times10^{6}M_{\odot}$ (the Freccia Rossa). If the $V_{\rm LSR}\approx-70$\,km\,s$^{-1}$ systemic velocity of CO-0.40-0.22 is consistent with that of the Freccia Rossa cloud, we hypothesise that the system is the remnant of an HVC that plunged into the Milky Way from the Galactic halo. This scenario may account for the kinematics of the CO-0.40-0.22 molecular cloud, leaving no extrinsic reason to invoke an associated IMBH.

\section*{Acknowledgements}

We thank the staff of CSIRO Astronomy and Space Science for the rapid scheduling of our observations. The Australia Telescope Compact Array is part of the Australia Telescope National Facility which is funded by the Australian Government for operation as a National Facility managed by CSIRO. ESP's research was funded in part by the Gordon and Betty Moore Foundation through grant GBMF5076. This work is based in part on observations made with the Spitzer Space Telescope, which is operated by the Jet Propulsion Laboratory, California Institute of Technology under a contract with NASA. This research has made use of the SIMBAD database, operated at CDS, Strasbourg, France.

\bibliographystyle{mnras}
\bibliography{cloud} 

\bsp	
\label{lastpage}
\end{document}